\documentclass[aps,pre,showpacs,notitlepage]{revtex4-1}
\usepackage{graphicx}
\usepackage{float}
\usepackage{amssymb,amsmath}
\usepackage{color}
\usepackage{bm}
\usepackage{subfigure}
\usepackage{ulem} %\sout{text}
\graphicspath{{./images/}}

\begin{document}

\normalem %this switches all the \emph in the text to italics, overriding the default underlining set in the ULEM package that is now present here.

\title{Absorbing and Shattered Fragmentation Transitions in Multilayer Coevolution}
\author{Marina Diakonova}
\author{Maxi San Miguel}
\author{V\'ictor M. Egu\'iluz}
\affiliation{Instituto de F\'isica Interdisciplinar y Sistemas Complejos IFISC (CSIC-UIB), E07122 Palma de Mallows, Spain}

%%%%%%%%%%%%%%%%%%%%%%%%%%%%%%%%%%%%%%%%%%%%%%%%%%%%%%%%%%%%%%%%%%%%%%%%%%%%%%%%%%%%%%%%%%%%%
\begin{abstract}
We introduce a coevolution voter model in a multilayer, by coupling a fraction of nodes across two network layers and allowing each layer to evolve according to its own topological temporal scale. When these time scales are the same the dynamics preserve the absorbing-fragmentation transition observed in a monolayer network at a critical value of the temporal scale that depends on interlayer connectivity. The time evolution equations obtained by pair approximation can be mapped to a coevolution voter model in a single layer with an effective average degree. When the two layers have different topological time scales we find an anomalous transition, named shattered fragmentation, in which the network in one layer splits into two large components in opposite states and a multiplicity of isolated nodes. We identify the growth of the number of components as a signature of this anomalous transition. We also find a critical level of interlayer coupling needed to prevent the fragmentation in a layer connected to a layer that does not fragment.
\end{abstract}

\maketitle
\section{Introduction} % (fold)
\label{sec:introduction}
The general framework linking together networks that represent different processes is that of a multilayer system \cite{Porter2013,Kivela2013}. Its significance has recently been highlighted in situations ranging from infrastructure \cite{Buldyrev2010}, information transmission and epidemic spreading \cite{Granell2013,Radicchi2013}, to social ties \cite{Szell2010,Halu2013} and others \cite{Gomez2013,Wang2013}. Dynamics in multilayer networks have been so far mostly analyzed in situations in which each layer is a fixed network. But even for solitary networks, the nodes state can evolve while the network itself is changing dynamically, an aspect that still needs to be incorporated into the multilayer framework \cite{Shai2013}. In particular we address here \emph{coevolution} dynamics, that is, coupled dynamics of node states and network topology in which the structure of the network becomes a variable \cite{Zimmermann2001,Zimmermann2004,Vazquez2008,Gross2008,Castellano2009}. This brings together dynamics \emph{of} the network with dynamics \emph{on} the network, going beyond situations of temporal networks decoupled from node state dynamics.

Coevolution dynamics in a single layer network has been considered in a variety of contexts \cite{Gross2008,Sayama2014} including social differentiation \cite{Eguiluz2005}, neural systems \cite{Meisel2012}, epidemic spreading \cite{Gross2006}, opinion formation \cite{Vazquez2008,Couzin2011,Zschaler2012}, cultural dynamics \cite{Vazquez2007,Centola2007} and ecosystems \cite{Dieckmann1999,Cantor2013}. The rewiring (plasticity) parameter $p$, measuring the relative time scale of evolution of the network and the states of the nodes, is typically the control parameter of coevolution dynamics. There a generic phenomenon is a fragmentation transition \cite{Vazquez2007,Vazquez2008} that splits the network into disconnected components. This transition occurs at a critical value $p_c$ of the rewiring parameter.

The Coevolving Voter Model (CVM)\cite{Vazquez2008} is an archetypal example displaying the fragmentation transition. The state of the system for coevolving networks is characterized through the interface density $\rho$ quantifying the fraction of edges linking nodes with different states (active links). When $\rho \neq 0$ the system is active, while for $\rho =0$ it is frozen, which in finite-size systems happens at finite times. In complex networks and if the rewiring probability $p$ is low enough, a single realization $\rho$ fluctuates around an asymptotic value $\rho^{\text{asym}}$, measured as the $t \rightarrow \infty$ limit of the interface density averaged over active runs at time $t$, $\rho^{\text{surv}}(t)$ \cite{Vazquez2008}. For $N \rightarrow \infty$ an absorbing transition from an active ($\rho^{\text{asym}}\neq 0)$ to a frozen ($\rho^{\text{asym}}=0$) state occurs at $p = p_c$. This transition coincides for finite-size systems with a fragmentation transition of the network freezing into two disconnected components for $p>p_c$, each one fully-ordered in one of the two possible states. The absorbing transition can be identified using $\rho$, while the fragmentation transition is identified by the relative size of the largest network component $S_1$ in the frozen state.

As a prototype situation to describe coevolution dynamics and fragmentation transitions in a multilayer we consider two coupled layers with Coevolving Voter Models. Each layer describes changes of state (for example opinion) by interactions in a given context with a different topological timescale as characterized by the rewiring parameter of the layer \cite{Csermely2013}. A key feature of our study is the flexibility of the strength of interlayer connectivity \cite{Buono2013,Shai2013}. This allows for the existence of nodes present in the two layers as well as other nodes only present in one of the layers. We call $q$, that varies between $0$ and $1$, the \emph{strength of multiplexing}: when $q$ is zero, the system consists of two fully-disconnected layers, whereas when $q$ is equal to unity we have a complete multiplex where all nodes exist in both layers.
% section introduction (end)

%%%%%%%%%%%%%%%%%%%%%%%
\section{The Model} % (fold)
\label{sec:the_model}
We couple together two binary state CVM into a multilayer system. Each layer $l\in{1,2}$ contains a network with $N_l$ nodes and an average degree $\mu_l$, where the state of each node can be $\pm 1$. In order to compare with previous results in monolayer networks, $N_1 = N_2 = N$ and layers are degree regular random networks (no self-loops are allowed) with $\mu_1 = \mu_2  = \mu = 4$. Initial states of the network nodes are random and equiprobable. We link the two layers by identifying a proportion $q$ of nodes across layers. Each CVM is characterized by its rewiring $p_1$ and $p_2$. There are three key parameters in our model: the plasticity of each layer (given by $p_1$ and $p_2$) and the strength of multiplexing $q$. 

A timestep is defined by $N$ updates, where each update involves selecting a random layer and evolving it with CVM rules. Since we require that nodes connected across layers are the same, any change in their states instantly propagates across the layers and changes their interlayer counterpart. To evolve a single CVM network a node $i$ in that layer $l$ is randomly selected. Its state is compared to that of a randomly chosen neighbor $j$ (in the same layer) and:
\begin{enumerate}
 \item nothing happens if the two are the same;
 \item otherwise, with probability $1 - p_l$, node $i$ copies the state of $j$, or else (with probability $p_l$) it severs the connection with $j$ and draws a link to a node randomly chosen from the set of nodes in layer $l$ that have the same state as $i$ but are not connected to it (if the set is empty, no rewiring is made).
\end{enumerate}

%%%%%%%%%%%%%%%%%%%%%%%%%%%%%%%%%%%%%%%%%%Fig. 1
\begin{figure} %[h!]
\centering
% \hspace*{-0.1 in}
\includegraphics[width = 0.5\textwidth]{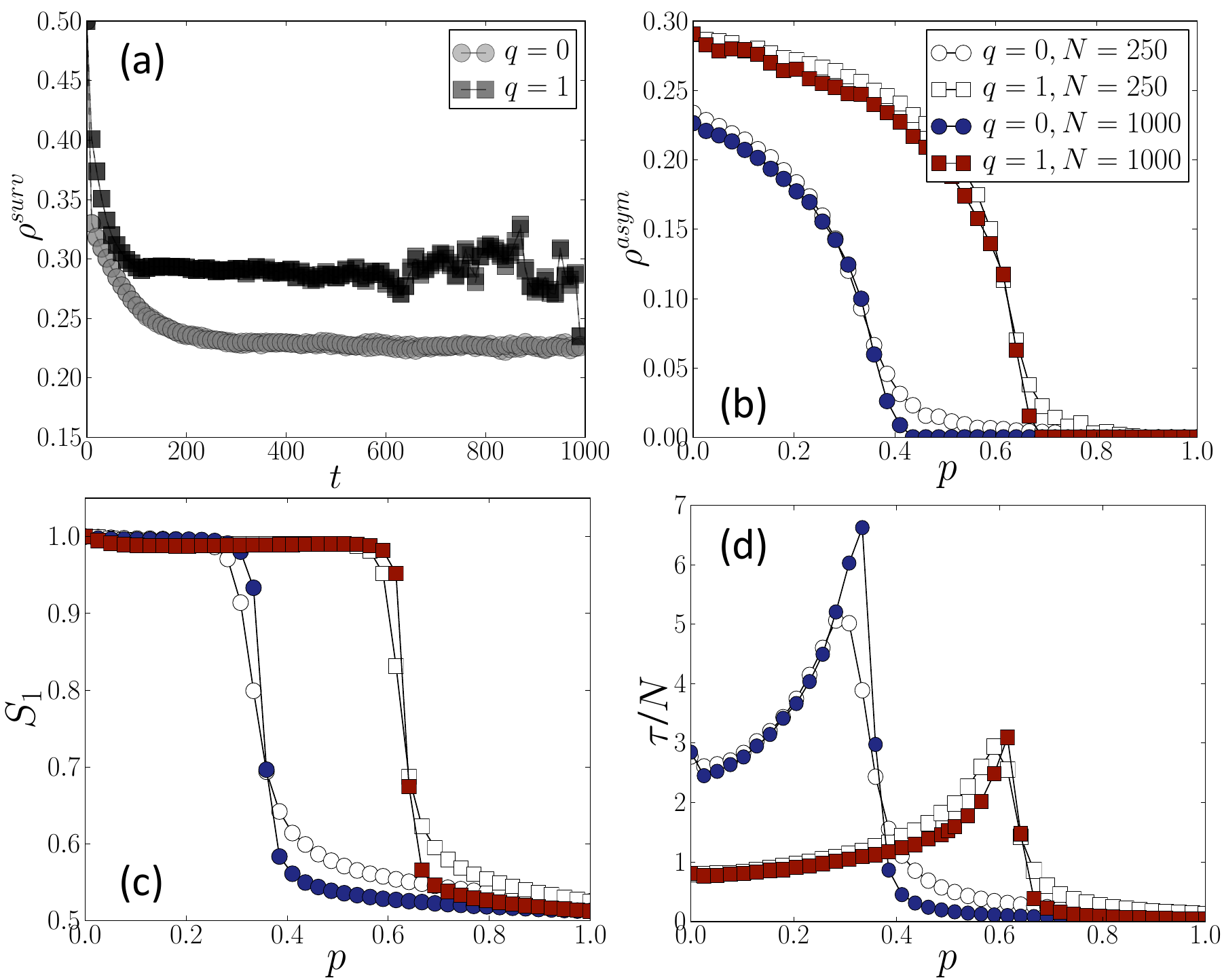}
\caption[]{(a) Interface density in a multilayer network of $N = N_1 = N_2 = 250$ nodes, averaged over realizations still active at time $t$, for $p = p_1 = p_2 = 0$ and interlayer connectivity $q$. The plateau indicates the asymptotic interface density shown in panel (b). (c) Average size of the largest cluster in the frozen state. (d) Scaled average time to reach an absorbing state. In all setups the ensemble consists of $10^4$ realizations. Parameter values for (b-d) are shown in panel (b).}
\label{transitions}
\end{figure}
% section the_model (end)
%%%%%%%%%%%%%%%%%%%%%%%%%
\section{Transition in a Symmetric Multilayer} % (fold)
\label{sec:transition_in_a_symmetric_multilayer}
The \emph{symmetric} multilayer system corresponds to $p = p_1 = p_2$. Statistical equivalence of initial conditions for each layer means average variables {\it e.g.}, the density of interfaces, {\it etc.} show the same behavior. For the voter ($p = 0$, \cite{Liggett1975}) multilayer the main result is that stronger interlayer connectivity leads to a higher $\rho^{\text{asym}}$ (see Fig.~\ref{transitions}(a)). Thus multiplexing increases the fraction of active links, {\it i.e.}, the degree of disorder in the system.

The variation of $\rho^{\text{asym}}$ with $p$ is shown in Fig.~\ref{transitions}(b). For $q = 0$ we recover the absorbing transition of \cite{Vazquez2008}. It continues to exist as two identical layers are interconnected, but now the critical rewiring $p_c(q)$ shifts to larger values. This implies that there is a range of rewiring at which disconnected layers would freeze, but where any interlayer connection would keep the system active. The strength of multiplexing necessary to achieve this increases monotonically with $p$. This range is finite, {\it i.e.}, $p_c(1) < 1$: if the timescale on which the topology of the system changes is sufficiently large (large $p$), even a fully-connected multiplex will freeze.

%%%%%%%%%%%%%%%%%%%%%%%%%%%%%%%%%%%%%%%%%%%%%%%Fig. 2
\begin{figure}
%\centering
\hspace*{-0.2 in}
\subfigure[Stable internal solution]{
\includegraphics[width = 0.30\textwidth, keepaspectratio = true]{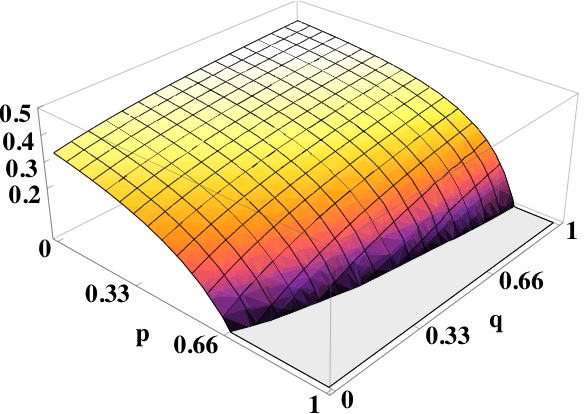}
\label{stab00:0}
}
\hspace*{-0.2 in}
\subfigure[Layer activity in $(p_1,p_2,q)$]{
\includegraphics[width = 0.40\textwidth, keepaspectratio = true]{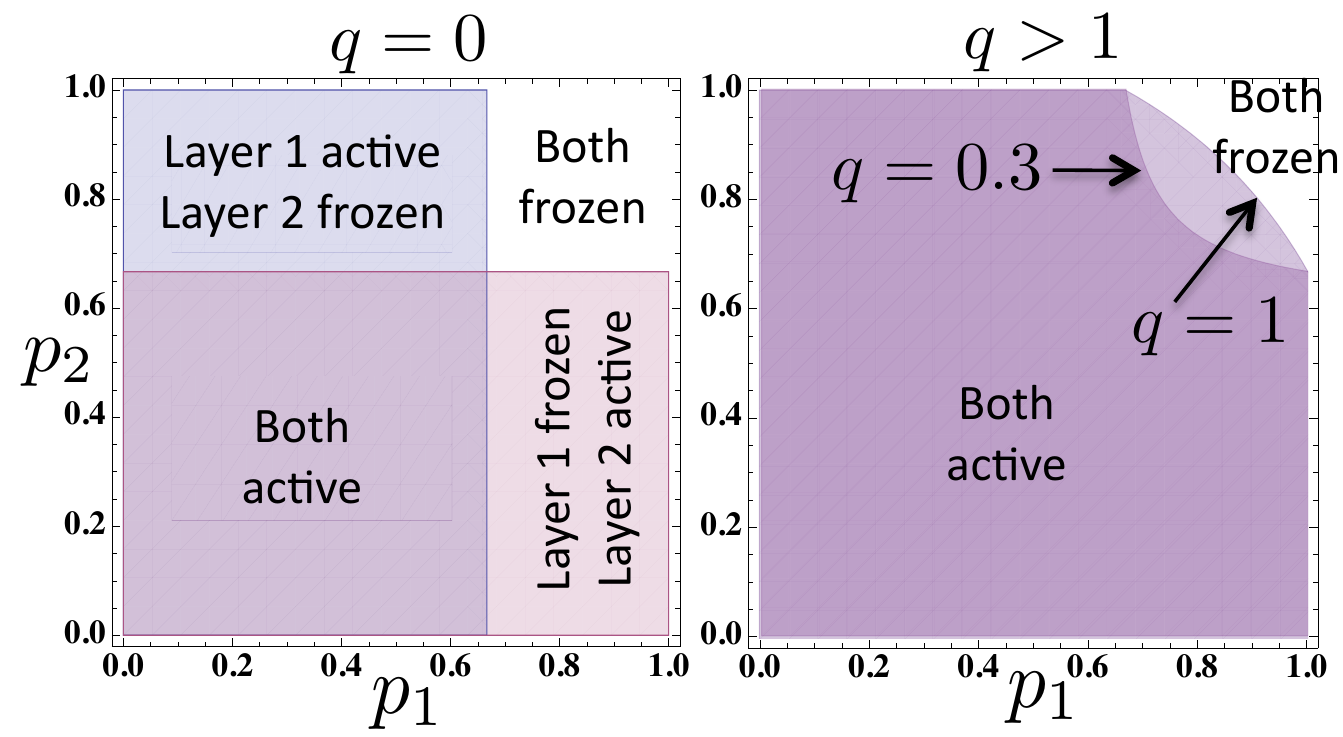}
\label{stab00:a}
}
\caption[]{(a) Interface density $\rho^{*}$  for the symmetric multiplex with $p = p_1 = p_2$ (given by Eq.~(\ref{eq:rhostar})). (b) Stability analysis of the $(\rho_1, \rho_2)=(0,0)$ fixed point. For $q>0$ both layers are active when $(0,0)$ is either unstable or saddle; system is frozen when $(0,0)$ is stable.}
\label{stab00}
%\vspace*{-0.2 in}
\end{figure}
The shift in the absorbing transition is mirrored by the offset in the fragmentation transition (Fig.~\ref{transitions}(c)): the multiplex can sustain a higher rate of rewiring with each layer still freezing into only one connected component. Above $p_c(q)$ each layer in the stationary system consists of two components with differing states, each such component connected to its counterpart in another layer. The characteristic times $\tau$, {\it i.e.}, the average time at which the multilayer system reaches an absorbing state (Fig.~\ref{transitions}(d)), diverges around $p_c(q)$ in a critical slowing down once more indicative of a fragmentation transition.
% section transition_in_a_symmetric_multilayer (end)

%%%%%%%%%%%%
\section{Pair-approximation for Interface Densities} % (fold)
\label{sec:pair_approximation_for_interface_densities}
We use a mean-field pair approximation \cite{Vazquez2008,Vazquez2008b,Vazquez2014} scheme to derive equations governing the evolution of interface densities $(\rho_1, \rho_2)$ of the two layers in the $N \rightarrow \infty$ limit. The equations for $l,m \in \{1,2\}$, $m \neq l$ are
\begin{equation}
\dot{\rho}_l = -2A_l \rho^2_l + \left( A_l - \frac{1}{\mu_l}\right) \rho_l + B_m \rho_m (1 - 2 \rho_l )~,
\label{eq:mf}
\end{equation}
where $A_l = \frac{1}{\mu_l}(1 - p_l)(\mu_l - 1)$, and $B_l = q(1 - p_l)$ quantifies the effect of the other layer on the interface density (note that there is no interlayer influence for either $q=0$ or only rewiring $p_l=1$). For the symmetric case ($p_1=p_2=p$) Eq.~(\ref{eq:mf}) can be written as the time evolution of a CVM in a monolayer with an effective average degree $\mu_\text{eff} = \mu(1+q)$. For the multilayer voter model ($p=0$) the stationary interface density is given by
\begin{equation}
\rho^*= \frac{\mu_\text{eff} -2}{2(\mu_\text{eff} - 1)}=\frac{\mu(1+q) -2}{2(\mu(1+q) - 1)}~,
\label{eq:rhostar}
\end{equation}
in agreement with the numerical simulations (Fig.~\ref{transitions}a) \cite{rho}. For arbitrary $p$ the stationary interface density surface $\rho^{*}(p,q) = \rho^{*}_1 = \rho^{*}_2$ is illustrated in Fig. ~\ref{stab00:0}. The system displays a phase transition at a critical rewiring probability $p_c(q)$ that depends on the degree of multiplexing $q$ in agreement with the numerical results (Fig.~\ref{transitions}b).

For arbitrary $(p_1,p_2)$ there are four fixed point solutions $(\rho^{*}_1, \rho^{*}_2)$, including $(0,0)$, with at most one such that $(\rho^{*}_1>0, \rho^{*}_2>0)$. If such a solution exists it is stable, and corresponds to a fully-active system with finite interface densities. In this case the $(0,0)$ origin is necessarily unstable, and hence the existence of an active state can be checked by examining the stability of $(0,0)$. Three parameter cross-sections at three different $q$ values are shown in Fig. \ref{stab00:a}. The multilayer dynamics exhibits three phases: both layers active, both frozen, and a mixed phase when one layer is active but the other is not. This latter phase exists only for a completely disconnected system, so that any degree of multiplexing ($q \neq 0$) is enough to tie the fate of one layer to that of another. In Fig.~\ref{stab00:a} the stable regime thus corresponds to the both-layers-frozen phase while the both-layers-active phase happens for both the unstable and saddle $(0,0)$, as long as the layers are connected.
In the asymmetric case ($p_1 \neq p_2$) the stable internal fixed point need not be located on the diagonal, meaning that the activity of the two layers need not be equal.
% section pair_approximation_for_interface_densities (end)
%%%%%%%%%%%%%%%%%%%%%%%%%%%%
\section{Anomalous Shattered Fragmentation in the Asymmetric Multiplex} % (fold)
\begin{figure} %[h!]
\centering
%\hspace*{-0.1 in}
\includegraphics[width = 0.5\textwidth, keepaspectratio = true]{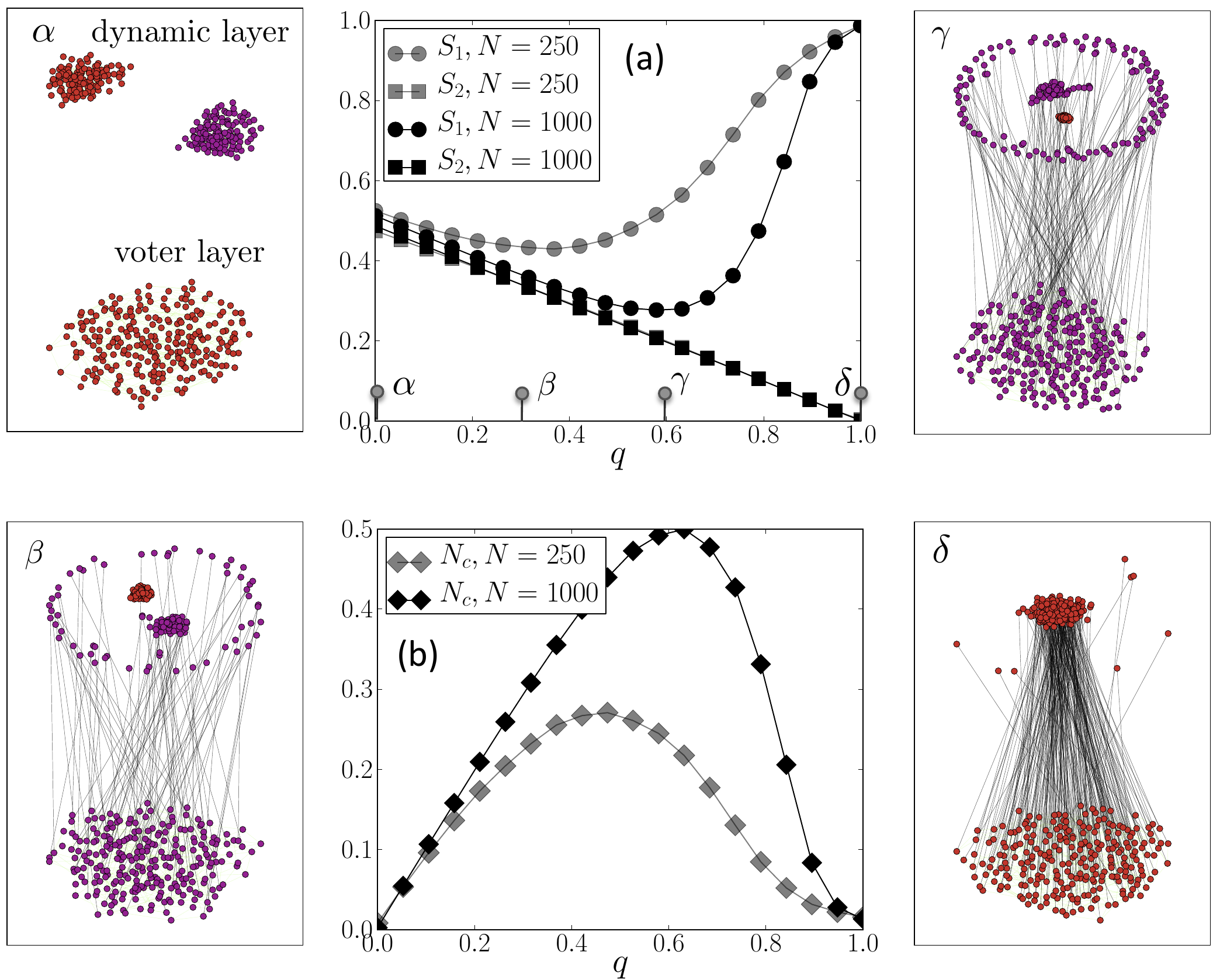}
\caption[]{(a) Relative size of $i^{\text{th}}$ largest components $S_i$ in the dynamic layer ($p_1 = 1$) coupled with strength $q$ to the voter layer ($p_2 = 0$), for systems with $N$ nodes in each layer. (b) Relative number of connected components $N_c$ in the dynamic layer. Variables are averaged over $10^4$ frozen configurations. Sides: typical snapshots of the $N = 250$ system for four sample values of $q$.}
\label{extreme_asymmetry}
\end{figure}

\label{sec:anomalous_shattered_fragmentation_in_the_asymmetric_multiplex}
The extreme asymmetry scenario couples a layer that only changes states ($p = 0$, which we call the \emph{voter} layer) and a layer that only rewires ($p = 1$, the \emph{dynamic} layer). Hence the voter layer is not affected by the dynamic layer. Instead it acts as a driver of the other layer, and thus does not fragment for any $q$ \cite{anom}.
For intermediary multiplexing, the dynamic layer displays an explosion in the number of disconnected nodes as a precursor of a anomalous fragmentation transition that we call \emph{shattered fragmentation} (Fig.~\ref{extreme_asymmetry}). Only two network components are ever significant in that layer, the rest being isolated nodes. For increasing $q$ the dynamic layer shows: (i) an increasing number of isolated components (which correspond to almost $qN$ nodes connected to the voter layer, Fig.~\ref{extreme_asymmetry}b), (ii) the second largest component composed of nodes disconnected from the voter layer and that were initially in the state opposite to that reached by the voter layer, thus, {\it i.e.}, $S_2=1/2 (1 - q)$, and (iii) the largest component formed by the remaining nodes. For larger $q$ its size $S_1$ increases until there is only one connected component left in that layer, which happens at $q = 1$.

%%%%%%%%%%%%%%%%%%%%%%%%%%%%%%%%%%%%%%%%%%%%%%%%%%%%%%%%%%%%%%Fig. 3

In the limit of infinite system size $S_1$ tails $S_2$ for a longer region of $q$, until $q^{*}$ defined as the minimum interlayer connectivity that realizes $S_1(q^*) = 1$ where $S_1(q)=\lim_{N \rightarrow \infty} S_1(q,N)$. We identify $q^{*}$ with a critical degree of multiplexing, or the minimum interlayer connectivity necessary to stop the dynamic layer from fragmenting. For extreme asymmetry, $q^{*}(p_1 = 1, p_2=0) = 1$, meaning that as long as $q<1$ it is impossible to prevent fragmentation of an infinitely large system for these parameters.

\begin{figure}[]
\centering
\hspace*{-0.1 in}
\includegraphics[width = 0.4\textwidth, keepaspectratio = true]{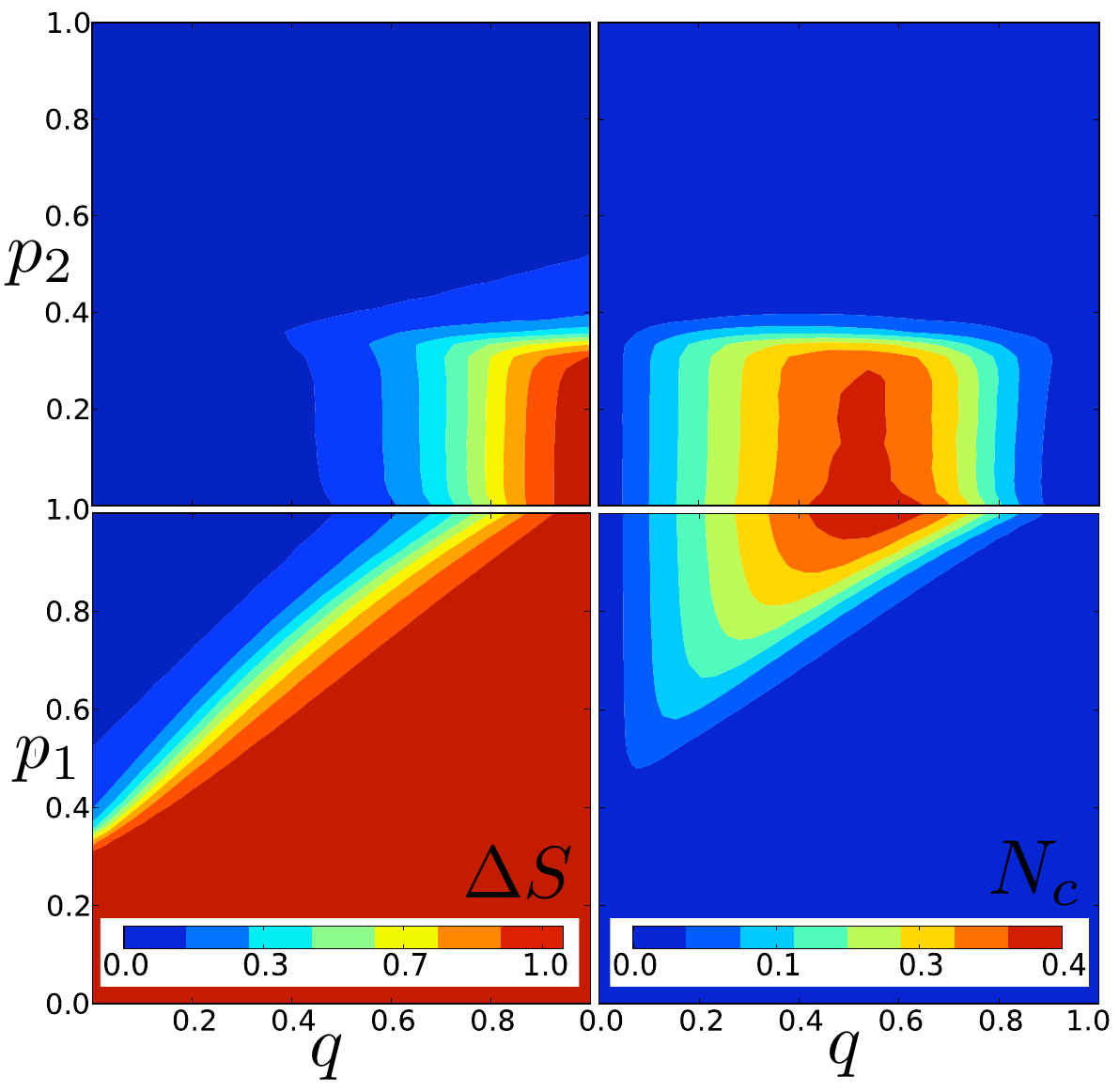}%general_fragmentation}
\caption[]{Shattered fragmentation of layer 1 in terms of the difference in relative size of its two largest components $\Delta S = S_1 - S_2$ (first column) and the relative number of connected components $N_c$ (second column), with $p_1 = 1$ (top row) and $p_2 = 0$ (bottom row). An unfragmented layer is red in $\Delta S$. Standard two-component fragmentation (blue $N_c$) becomes shattered as the number of isolated nodes increases (any deviation from blue in $N_c$). Quantities are averaged over $10^4$ realizations of the system with $N = 500$ nodes in each layer. Higher system size would have sharper transitions in $\Delta S$ and higher peaks in $N_c$ that also move to higher values of $q$.}
\label{general_fragmentation}
\end{figure}

\begin{figure}
\centering
\includegraphics[width = 1\textwidth, keepaspectratio = true]{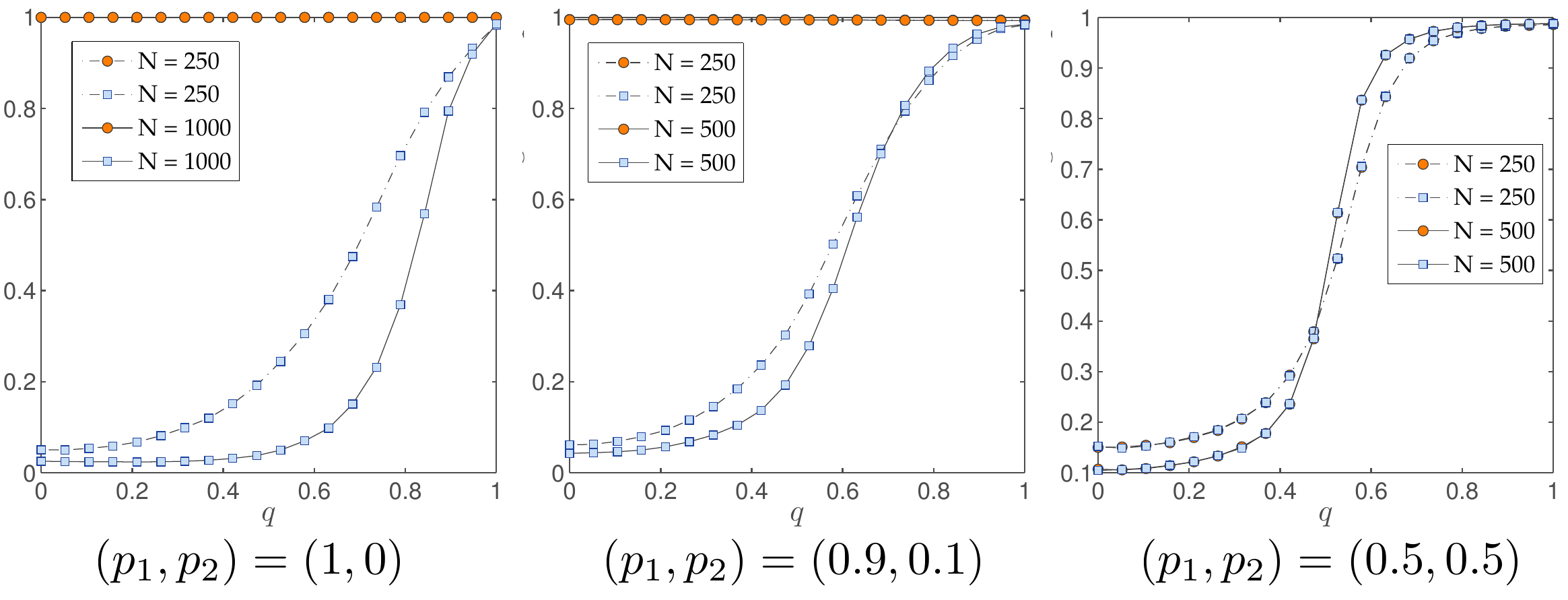}
\caption[]{Difference $\Delta S$ between two largest clusters in each layer (layer 1 in blue, layer 2 in orange), for different sample sizes $N$ and rewiring probabilities $(p_1,p_2)$, as a function of interlayer connectivity $q$. Cluster sizes are averaged over $10^4$ configurations of systems upon reaching the frozen state.}
\label{step_deltaS}
\end{figure}

\begin{figure}
\centering
\includegraphics[width = 1\textwidth, keepaspectratio = true]{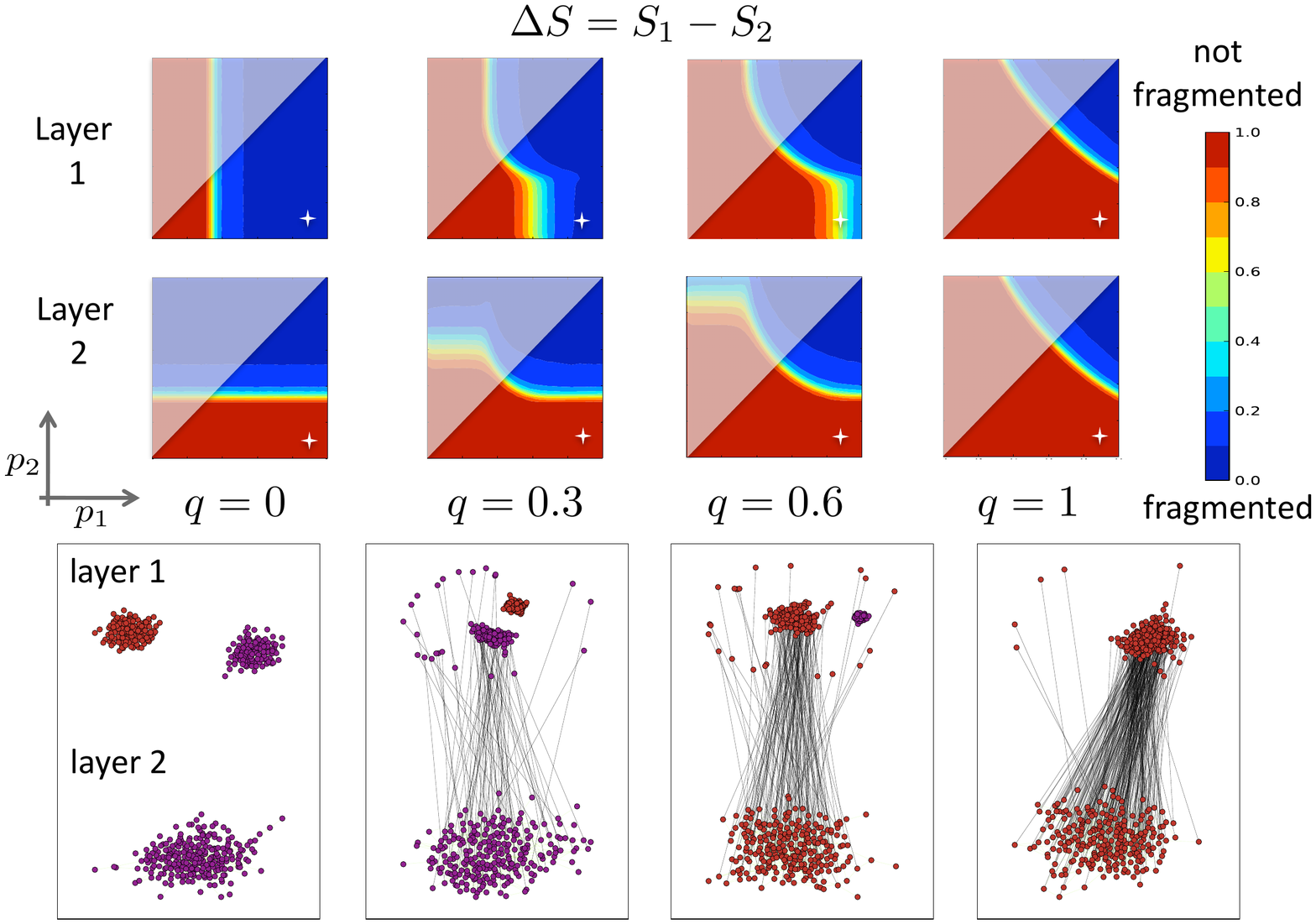}
\caption[]{Top two rows: difference $\Delta S$ between the two largest clusters in each layer, for a layer (top row corresponds to layer 1, second row to layer 2). Due to the symmetry in rewiring rates the second row is a symmetric transformation of the first, though for ease of reading both are shown alongside, and only the bright bottom triangles should be looked at. The squares correspond to $(p_1,p_2)$ cross-sections of the parameter hypercube at various $q$ values. Each point is an average over $10^4$ realisations of a multiplex with $N = 500$ nodes on each layer. The colour scheme ranges between the prevalent colours observed, with blue corresponding to $0$ and red to $1$. Bottom row: snapshot of a typical absorbing state computed at the starred point in the above rows, $(p_1,p_2) = (0.9, 0.1)$. See \textit{http://ifisc.uib-csic.es/lasagne/media/MultilayerNetworkMovies-q@.mp4}, where $@$ can be $0$, $0.5$ and $1$, for the dynamics behind the fragmentation of $(p_1,p_2) = (0.9, 0.1)$, $N = 250$.}
\label{FinalStateTopology}
\end{figure}

Shattered fragmentation is a general consequence of the rewiring asymmetry $p_1 \neq p_2$. Figure~\ref{general_fragmentation} quantifies it in terms of $\Delta S=S_1-S_2$ and $N_c$: $\Delta S$ informs on the existence of a fragmentation transition, and $N_c$, on the nature of fragmentation. It illustrates two ways of varying asymmetry: lowering/raising the rewiring of one layer keeping the other value fixed. As long as the states of the nodes of the more dynamic layer are allowed to change ($p \neq 1$), its fragmentation can be prevented by coupling it to a layer that, uncoupled, would not fragment. This corresponds to a master-slave coupling.
For any $(p_1,p_2)$, $\Delta S(q)$ displays a steeper transition for increasing $N$, suggesting a step transition in the thermodynamic limit (see Fig.~\ref{step_deltaS}). The fact that fragmentation in one layer need not necessarily entail fragmentation in another, is not a feature of extreme asymmetry (Fig.~\ref{FinalStateTopology}). The critical degree of multiplexing decreases from extreme asymmetry $q^{*}(p_1 = 1, p_2=0) = 1$ to symmetry $q^{*}(p_1 = 0.5, p_2=0.5) = 0.5$ Fig.~ \ref{step_deltaS}, lowering for smaller rewiring. When the (more) static layer is the voter model, $q^{*}(p_1 > p_c(0))$ follows a quasi-linear dependence on $p_1$. On the other hand, for $N \rightarrow \infty$, $q^{*}(p_1=1, p_2)\rightarrow 1$. Fragmentation of the (more) dynamic layer is maximized by asymmetry where it shatters the layer into isolated components. Their fraction decreases as the layer becomes more static Fig.~(\ref{general_fragmentation}D), but is not dependent on the exact extent of rewiring of the stabilizing layer Fig.~(\ref{general_fragmentation}B).

Finally, we note that the absorbing and fragmentation transitions coincide in simulations of even the asymmetric multiplex, just as in solitary CVM. Therefore we associate $q^{*}$ with the minimal multiplexing necessary to keep the more dynamic layer active. The pair approximation fails to capture the anomalous transition since the nature of shattered fragmentation is indicative of the presence of isolated nodes. Thus analytics suggest that even small degree of multiplexing is sufficient to keep the dynamic level active, and do not take into account that the flux of links away from the interconnected nodes might still keep it topologically frozen.

%%%%%%%%%%%%%%%%%%%%%%%%%%%%%%%%%%%%%%%%%%%%%%%%%%%%%%%%%%%%%%%%%%%%%%%%%%%%%Fig 4
% section anomalous_shattered_fragmentation_in_the_asymmetric_multiplex (end)
%%%%%%%%%%%%%%%%%%%%%%%
\section{Summary and conclusions} % (fold)
\label{sec:summary_and_conclusions}
We have analyzed a multilayer system constructed by coupling together with an arbitrary degree of multiplexing $q$ two coevolving networks with different rewiring parameters. The multilayer structure offsets the critical value of the rewiring for the occurrence of absorbing and fragmentation transitions, hence multiplexing is shown to be able to prevent network fragmentation. We have also found a critical degree of multiplexing characterized as the minimal required interlayer connectivity necessary to stop the fragmentation of a layer by coupling it to a layer that does not fragment. This critical value is a function of the rewiring parameters of the two layers. Subcritical multiplexing leads to the existence of a shattered fragmentation typical of the more topologically dynamic layer as a consequence of dynamic asymmetry between the layers. This phenomenon, in which network fragmentation results in an explosion of isolated nodes as the strength of multiplexing approaches its critical value, is not captured by a pair-approximation calculation. Other approaches should be explored to handle analytically the growth of isolated nodes \cite{Vazquez2014}.
% section summary_and_conclusions (end)

\label{sec:supplementary material}

This work has been supported by the Spanish MINECO and FEDER under projects INTENSE@COSYP (FIS2012-30634) and MODASS (FIS2011-24785), and by the EU Commission through the project LASAGNE (FP7-ICT-318132).

\bibliography{CVM_bibliography}

\begin{thebibliography}{32}
\bibitem [{ano()}]{anom}%
  \BibitemOpen
  \href@noop {} {\ }\bibinfo {note} {As long as one of the layers only rewires
  $p=1$, the other layer behaves as if it were disconnected and thus fragments
  at $p_c(q=0)$.}\BibitemShut {Stop}%
\end{thebibliography}

% \begin{thebibliography}{20}
% \end{thebibliography}

\end{document}